\documentclass[aps,prd,amsmath,amssymb,reprint,author-numerical,onecolumn]{revtex4-1}

\begin{document}

\title{Cho-Duan-Ge decomposition of QCD in the constraintless Clairaut-type formalism}
\author{Michael L. Walker}
\affiliation{University of Melbourne, Parkville, VIC 3010, Australia}
\author{Steven Duplij}
\affiliation{Universit\"{a}t M\"{u}nster, Einsteinstr. 62, 48149 M\"{u}nster, Germany
}

\begin{abstract}
We apply the recently derived constraintless 
Clairaut-type formalism to the Cho-Duan-Ge 
decomposition in SU(2) QCD. 
We find nontrivial corrections to the 
physical equations of motion and that the contribution 
of the topological degrees of freedom is qualitatively 
different from that found by treating the monopole potential 
as though it were dynamic. We also find alterations to the 
field commutation relations that undermine the particle 
interpretation in the presence of the chromomonopole condensate.
\end{abstract}
\maketitle

\section{Introduction}

\label{sec:intro} The occurence of \textquotedblleft
redundant\textquotedblright\ degrees of freedom not determined by
equations of motion (EOMs) is a characteristic property of any physical system
having symmetry \cite{sundermeyer,reg/tei}. In gauge theories the covariance 
of EOM under symmetry transformations leads to gauge ambiguity,
\emph{i.e.} the appearence of undetermined functions.
In this situation some dynamical
variables obey first order differential equations \cite{hen/tei}.
One then employs
a suitably modified Hamiltonian formalism, such as the Dirac 
theory of constraints \cite{dirac}.

A constraintless generalization of the Hamiltonian formalism based on a
Clairaut-type formulation was recently put forward by one of the authors
\cite{D10,D14}. It generalises the standard Hamiltonian formalism to include
Hessians with zero determinant, providing
a rigorous treatment of the non-physical degrees of freedom in the derivation
of EOMs and the quantum commutation relations. An outline is given in
app.~\ref{app:Clairaut}.

The Cho-Duan-Ge (CDG) decomposition of the gluon field in Quantum
Chromodynamics (QCD), published by Duan and Ge \cite{DG79} and also by Cho
\cite{Cho80a}, specifies the Abelian components of the background field in a
gauge covariant manner. In so doing it identifies the monopole degrees of 
freedom (DOFs)
of the gluon field naturally, making it preferable to the conventional maximal Abelian
gauge \cite{tH81}. It can also generate a gauge invariant canonical momentum,
which makes it of interest to studies of nucleon spin decomposition
\cite{W11,CSWG11,ZP12,CGZ12}.

Up until now, the monopole DOFs have not been rigorously handled. Indeed,
merely accounting for the physical and gauge DOFs proved to be a long and
difficult task \cite{S99,CP02,KMS05,K06,BCK02}. An important observation of
the monopole DOFs by  Cho \emph{et al.} is that the Euler-Lagrange equation
for the Abelian direction does not yield a new EOM. Their interpretation is
that the
monopole is the ``slow-changing background part'' of the gauge field while the
physical gluons constituted the ``fast-changing quantum part''.

In this paper we apply the Clairaut formalism to the monopole DOFs in
two-colour QCD. We consider both the gluon field and scalar ``quarks'' in the
fundamental field. We find that the interaction between monopole and physical
DOFs vanishes from the EOMs, but that the canonical commutation relations are
altered in a manner that leaves particle number undefined.

Section \ref{sec:CDG} describes the CDG decomposition and establishes
notation. In section \ref{sec:qalpha} we identify the field theory equivalent
of $q^{\alpha}$ and go on to find the $q^{\alpha}$ curvature in section
\ref{sec:qfieldstrngh}. The curvature's non-zero value leads to alterations in
the EOMs elucidated in section \ref{sec:altEOM}, while corresponding results
are found in section \ref{sec:fundamental} for colour-charged scalars in the
fundamental representation. Our most important results, alterations to the
commutation relations and their implications for the particle interpetation,
are discussed in section \ref{sec:commutator}. We give a final discussion in
section \ref{sec:discuss} and a detailed summary of the Clairaut
formalism in app.~\ref{app:Clairaut}.

\section{\label{sec:CDG}Representing the Gluon Field}

The Cho-Duan-Ge (CDG) decomposition \cite{DG79,Cho80a}, and another like it
\cite{FN99c}, was (re-)discovered \cite{S99} at about the turn of the century
when several groups were readdressing the stability of the chromomonopole
condensate \cite{CP02,KMS05,K06,CmeP04,KKP05,me07}. Some authors
\cite{KMS05,K06,KKP05}, including one of the current ones \cite{me07}, have
overlooked the differences between the CDG decomposition and that of Faddeev
and Niemi, referring to the former as either the Cho-Faddeev-Niemi (CFN) or
the Cho-Faddeev-Niemi-Shabanov (CFNS) decomposition. In this paper we label it
the CDG decomposition, as per the convention of Cho \emph{et al.} \cite{CGZ12}.

The Lie group $SU(N)$ has $N^{2}-1$ generators $\lambda^{(a)}$ ($a=1,\ldots
N^{2}-1$), of which $N-1$ are Abelian generators $\Lambda^{(i)}$ ($i=1,\ldots
N-1$).
The gauge transformed Abelian directions (Cartan generators) are denoted as%
\begin{equation}
\hat{n}_{i}(x)=U(x)^{\dagger}\Lambda^{(i)}U(x).
\end{equation}

Gluon fluctuations in the $\hat{n}_{i}(x)$ directions are described by
$c_{\mu}^{(i)}(x)$, where $\mu$ is the Minkowski index. There is a covariant
derivative which leaves the $\hat{n}_{i}(x)$ invariant,
\begin{equation}
\label{eq:Dhat}\hat{D}_{\mu}\hat{n}_{i}(x)\equiv(\partial_{\mu}+g\vec{V}_{\mu
}(x)\times)\hat{n}_{i}(x)=0,
\end{equation}
where $\vec{V}_{\mu}(x)$ is of the form
\begin{equation}
\label{eq:vecV}\vec{V}_{\mu}(x)=c_{\mu}^{(i)}(x)\hat{n}_{i}(x)+\vec{C}_{\mu
}(x), \quad\vec{C}_{\mu}(x)=g^{-1}\partial_{\mu}\hat{n}_{i}(x)\times\hat
{n}_{i}(x).
\end{equation}
The vector notation refers to the internal space, and summation is implied
over $i=1,\ldots N-1$. For later convenience we define
\begin{equation}
F^{(i)}_{\mu\nu}(x) = \partial_{\mu}c^{(i)}_{\nu}(x)- \partial_{\nu}%
c^{(i)}_{\mu}(x)
\end{equation}

\begin{equation}
\vec{H}_{\mu\nu}(x) = \partial_{\mu}\vec{C}_{\nu}(x)- \partial_{\nu}\vec
{C}_{\mu}(x) +g\vec{C}_{\mu}(x)\times\vec{C}_{\nu}(x)= H^{(i)}_{\mu\nu}(x)
\hat{n}_{i}(x),
\end{equation}
\begin{equation}
H^{(i)}_{\mu\nu}(x) = \vec{H}_{\mu\nu}(x) \cdot\hat{n}_{i}(x).
\end{equation}

The vectors $\vec{X}_{\mu}(x)$ denote the dynamical components of the gluon
field in the off-diagonal directions of the internal space, so if $\vec
{A}_{\mu}(x)$ is the gluon field then%

\begin{equation}
\vec{A}_{\mu}(x)=\vec{V}_{\mu}(x)+\vec{X}_{\mu}(x)=c_{\mu}^{(i)}(x)\hat{n}%
_{i}(x) +\vec{C}_{\mu}(x)+\vec{X}_{\mu}(x),
\end{equation}
where
\begin{equation}
\vec{X}_{\mu}(x) \bot\hat{n}_{i}(x),\; \forall\, 1\le i<N\,,\quad\vec{D}_{\mu}
=\partial_{\mu}+g\vec{A}_{\mu}(x).
\end{equation}

The Lagrangian density is still
\begin{align} \label{eq:Lagrange}
\mathcal{L}_{gauge}(x) = -\frac{1}{4} \vec{F}_{\mu\nu}(x) \cdot\vec{F}^{\mu
\nu}(x)
\end{align}
where the field strength tensor of QCD expressed in terms of the CDG
decomposition is
\begin{align} \label{eq:field strength}
\vec{F}_{\mu\nu}(x)  &  =(F^{(i)}_{\mu\nu}(x) + H^{(i)}_{\mu\nu}(x)) \hat
{n}_{i}(x)
+(\hat{D}_{\mu}\vec{X}_{\nu}(x)-\hat{D}_{\nu}\vec{X}_{\mu}(x)) +g\vec
{X}_{\mu}(x)\times\vec{X}_{\nu}(x).
\end{align}

We will later have need of the conjugate momenta. These are only defined up to
a gauge transformation, so to avoid complications we take the Lorenz gauge.
The conjugate momentum for the Abelian component is then
\begin{equation}
\Pi^{(i)\mu}(x)= \frac{\delta \Big( \int d^3x \mathcal{L}_{gauge}\Big)}{\delta\partial_{0} c^{(i)}_{\mu}(x)} 
=-\vec{F}^{0\mu}(x) \cdot\hat{n}^{(i)}(x),
\end{equation}
while the conjugate momentum of $\vec{X}_{\mu}(x)$ is
\begin{equation} \label{eq:conjugateX}
\vec{\Pi}^{\mu}(x)= \frac{\delta \Big( \int d^3x \mathcal{L}_{gauge}\Big)}{\delta\hat{D}_{0} \vec{X}_{\mu}(x)}
=-\frac{1}{2} \Big(\hat{D}^{0} \vec{X}^{\mu}(x)-\hat{D}^{\mu}\vec{X}^{0}(x) +g(\vec{X}^\mu(x) \times \vec{X}^\nu (x))_{\perp \{\hat{n}^{(i)}:1\le i \le M\}} \Big).
\end{equation}
From now on we restrict ourselves to the $SU(2)$ theory, for which there is
only one $\hat{n}(x)$ lying in a three dimensional internal space, and neglect
the $(i)$ indices. The results can be extended to larger $SU(N=M+1)$ gauge 
groups \cite{me07}, although the cross-product in eq.~(\ref{eq:conjugateX})
vanishes when $N=2$.

The above outline neglects various mathematical subtleties involved in a fully
consistent application of the CDG decomposition. In fact, its proper
interpretation and gauge-fixing took considerable effort by several
independent groups. The interested reader is referred to
\cite{S99,CP02,KMS05,K06,BCK02} for further details.

\section{\label{sec:qalpha}The $q^{\alpha}$ gauge fields of the monopole
field}

Now we adapt the Clairaut approach (see app.~\ref{app:Clairaut}) 
\cite{D11,D10} to quantum field theory and
apply it to the CDG decomposition of the QCD gauge
field, leaving the fundamental representation until section
\ref{sec:fundamental}. Substituting the polar angles,
\begin{equation} \label{eq:npolar}
\hat{n}(x)=\cos\theta(x) \sin\phi(x) \,\hat{e}_{1} +\sin\theta(x) \sin
\phi(x)\,\hat{e}_{2} +\cos\phi(x)\,\hat{e}_{3}.
\end{equation}
and defining
\begin{align}   \label{eq:nphipolar}
\sin\phi(x) \,\hat{n}_{\theta}(x)  &  \equiv\int dy^{4} \frac{d\hat{n}%
(x)}{d\theta(y)} =\sin\phi(x)\,(-\sin\theta(x)\,\hat{e}_{1}+\cos
\theta(x)\,\hat{e}_{2})\nonumber\\
\hat{n}_{\phi}(x)  &  \equiv\int dy^{4} \frac{d\hat{n}(x)}{d\phi(y)}
=\cos\theta(x)\cos\phi(x)\,\hat{e}_{1}+\sin\theta(x)\cos\phi(x)\,\hat{e}%
_{2}-\sin\phi(x)\,\hat{e}_{3},
\end{align}
for later convenience, we note that 
\begin{equation} \label{eq:nphiphi}
\hat{n}_{\phi\phi} = -\hat{n},\; 
\sin \phi \,\hat{n}_{\theta\theta} = -\sin\phi(x)\, (\cos\theta \,\hat{e}_1 + \sin\theta\, \hat{e}_2),
\end{equation}
and that the vectors $\hat{n}(x)=\hat{n}_{\phi}(x)\times\hat{n}_{\theta}(x)$ 
form an orthonormal basis of the internal space.

Substituting the above into the Cho connection in eq.~(\ref{eq:vecV}) gives
\begin{align}
g\vec{C}_{\mu}(x)  &  =(\cos\theta(x) \cos\phi(x) \sin\phi(x)\partial_{\mu
}\theta(x) +\sin\theta(x)\partial\phi(x))\,\hat{e}_{1}\nonumber\\
& +(\sin\theta(x) \cos\phi(x) \sin\phi(x) \partial_{\mu}\theta(x) -\cos
\theta(x) \partial\phi(x))\,\hat{e}_{2} -\sin^{2}\phi(x)\partial_{\mu}%
\theta(x)\,\hat{e}_{3}\nonumber\\
&  =\sin\phi(x)\,\partial_{\mu}\theta(x)\, \hat{n}_{\phi}(x)-\partial_{\mu
}\phi(x)\,\hat{n}_{\theta}(x)
\end{align}
from which it follows that
\begin{equation}
g^{2}\vec{C}_{\mu}(x)\times\vec{C}_{\nu}(x) =\sin\phi(x)(\partial_{\mu}\phi(x)
\partial_{\nu}\theta(x) -\partial_{\nu}\phi(x)\partial_{\mu}\theta(x))\hat
{n}(x),\label{eq:CxC}%
\end{equation}

Treating $\theta,\phi$ as dynamic variables, their conjugate momenta are
\begin{align}
\bar{p}_{\phi}(x) & = \int dy^{3} \frac{\delta\mathcal{L}}%
{_{x}\partial_{0} \phi(x)}\nonumber\\
& =\int dy^{3} \int dy^{0} \delta(x^{0} - y^{0}) \Big( \sin\phi(y)
_{y}\partial^{\mu}\theta(y)\hat{n}(y) +\hat{n}_{\theta}(y)\times\vec{X}^{\mu
}(y)\Big)\cdot\vec{F}_{0\mu}(y) \delta^{3}(\vec{x}-\vec{y})\nonumber\\
& = \Big(\sin\phi(x) \partial^{\mu}\theta(x)\hat{n}(x) +\hat{n}_{\theta
}(x)\times\vec{X}^{\mu}(x)\Big)\cdot\vec{F}_{0\mu}(x),\label{eq:qphi}\\
 \bar{p}_{\theta}(x)&= \int dy^{3} \frac{\delta\mathcal{L}%
}{_{x}\partial_{0} \theta(x)}\nonumber\\
&  =-\int dy^{3} \int dy^{0} \delta(x^{0}-y^{0}) \sin\phi(y) \Big(
_{y}\partial^{\mu}\phi(y)\,\hat{n}(y) +\sin\phi(y)\,\hat{n}_{\phi}(y)\times
\vec{X}^{\mu}(y)\Big)\cdot\vec{F}_{0\mu}(y) \delta^{3}(\vec{x}-\vec{y})\nonumber\\
& =-\sin\phi(x) \Big(\partial^{\mu}\phi(x)\,\hat{n}(x)+ 
\hat{n}_{\phi}(x)\times\vec{X}^{\mu}(x)\Big)\cdot\vec{F}_{0\mu}(x) .
\label{eq:qtheta}%
\end{align}

The Hessian is given by
\begin{equation}
\left\Vert \frac{\delta^{2}\mathcal{L}}{\delta q^{A} \delta q^{B}}\right\Vert = 0,
\end{equation}
where $A, B$ run over all fields, both physical and topological. 
It follows from inspection of the Lagrangian density, eqs (\ref{eq:Lagrange}, \ref{eq:field strength}), 
that the time
derivatives of $\theta(x), \phi(x)$ occur only in linear combination with
those of one of the physical gluon fields $c_\mu(x), \vec{X}_\mu(x)$, either through
$F_{0\nu}(x) + H_{0\nu}(x)$ or $\hat{D}_0$. (This is readily extended to quarks, which we 
introduce in section \ref{sec:fundamental}). Therefore the rows (columns)
of the Hessian matrix corresponding to $\dot{\theta}(x), \dot{\phi}(x)$ must be linear
combinations of those corresponding to the physical field velocities, so the Hessian vanishes.

This linear dependence within the Hessian is consistent with Cho and Pak's \cite{CP02}, and Bae 
\emph{et al.}'s \cite{BCK02} finding that $\hat{n}(x)$ (and by extension 
$\theta(x),\phi(x)$) does not generate an independent EOM.

We therefore use the discussion surrounding
(3.10) in \cite{D10} and define
\begin{align}
B_{\theta}(x)  & \equiv \bar{p}_{\theta}(x),\;
B_{\phi}(x) \equiv \bar{p}_{\phi}(x) .
\end{align}
where the definitions of $B_{\phi}(x),B_{\theta}(x)$ are generalised to
quantum field theory from those in \cite{D10}. It follows that
$H_{phys}=H_{mix}$ (also defined in \cite{D10}). 


\section{The $q^{\alpha}$-curvature} \label{sec:qfieldstrngh}
From eqs.~(\ref{eq:qphi},\ref{eq:qtheta}) we have
\begin{align}
\frac{\delta B_{\phi}(x)}{\delta\theta(y)}=  
&  \Big( \sin\phi(x)\hat{n}_{\theta\theta}(x) \times \vec{X}^\mu \cdot \vec{F}_{0\mu}(x) - T_\phi (x) \Big)
\delta^{4}(x-y), \\
\frac{\delta B_{\theta}(x)}{\delta\phi(y)}=  
&  -\Big( \cos \phi(x)\Big(\partial^{\mu}\phi(x)\,\hat{n}(x)+\hat{n}_{\phi}(x)
\times\vec{X}^{\mu}(x)\Big)\cdot \Big( \vec{F}_{0\mu}(x) + \vec{H}_{0\mu} \Big) + T_\theta (x) \Big)
\delta^{4}(x-y),
\end{align}
where
\begin{align}
T_\phi (x) = & \partial^k \Big[ \sin \phi(x) \, \hat{n} \cdot \vec{F}_{0k} (x)
- \Big( \sin \phi(x) \partial_k \theta (x) 
+ \hat{n}_\theta (x) \times \vec{X}_k \cdot \hat{n} \Big) \partial_0 \phi(x) \Big],  \\
T_\theta (x) = & - \partial^k \Big[ \sin \phi(x) \Big( \hat{n} \cdot \vec{F}_{0k} (x) 
+ \Big( \partial_k \phi (x) + \hat{n}_\phi (x) \times \vec{X}_k \cdot \hat{n} \Big) \partial_0 \theta(x)
\Big)\Big],
\end{align}
are the surface terms arising from derivatives 
$\frac{\delta (\partial \theta)}{\delta \theta},\,\frac{\delta (\partial \phi)}{\delta \phi}$ and the latin index $k$
is used to indicate that only spacial indices are summed over.

This yields the $q^{\alpha}$-curvature
\begin{align}
\label{eq:curvature}\mathcal{F}_{\theta\phi}(x) =&\int dy^{4} \Big( \frac
{\delta B_{\theta}(x)}{\delta\phi(y)}-\frac{\delta B_{\phi}(x)}{\delta
\theta(y)}\Big)\delta^{4}(x-y) +\{B_{\phi}(x),B_{\theta}(x)\}_{phys}%
\nonumber\\
=&-\cos\phi(x)\Big(\partial^{\mu}\phi(x)\,\hat{n}(x)+ \hat{n}_{\phi}(x) \times
\vec{X}^{\mu}(x) \Big) \cdot \Big(\vec{F}_{0\mu}(x) + \vec{H}_{0\mu}(x) \Big)
-\sin\phi(x)\,\hat{n}_{\theta\theta}(x) \times \vec{X}^\mu(x) 
\cdot\vec{F}_{\mu 0}(x)  \nonumber \\
&  +T_\phi (x) - T_\theta (x).
\end{align}
where we have used that the bracket $\{B_{\phi}(x),B_{\theta}(x)\}_{phys} $
vanishes because $B_{\phi}(x)$ and $B_{\theta}(x)$ share the
same dependence on the dynamic DOFs and their derivatives.

In earlier work on the Clairaut formalism \cite{D11,D10} this was called the
$q^{\alpha}$-field strength, but we call it $q^{\alpha}$-curvature in quantum
field theory applications to avoid confusion.

This non-zero $\mathcal{F}^{\theta\phi}(x)$ is necessary, and usually
sufficient, to indicate a non-dynamic contribution to the conventional
Euler-Lagrange EOMs. More
significant is a corresponding alteration of the quantum commutators, with
repurcussions for canonical quantisation and the particle number.

\section{Altered equations of motion} \label{sec:altEOM}
Generalizing eqs.~(7.1,7.3,7.5) in \cite{D10},
\begin{equation}
\label{eq:alterEOM}\partial_{0}q(x)=\{q(x),H_{phys}\}_{new}= \frac{\delta
H_{phys}}{\delta p(x)} -\int dy^{4} \sum_{\alpha=\phi,\theta}
\frac{\delta B_{\alpha}(y)}{\delta p(x)} \partial^{0}\alpha(y),
\end{equation}
the derivative of the Abelian component, complete with corrections from the
monopole background is 
\begin{equation} \label{eq:altEOMc}
\partial_{0}c_{\sigma}(x) =\frac{\delta H_{phys}}{\delta\Pi^{\sigma}(x)}
-\int dy^{4} \sum_{\alpha=\phi,\theta} 
\frac{\delta B_{\alpha}(y)}{\delta\Pi^{\sigma}(x)}\partial^{0}\alpha(y).
\end{equation}

The effect of the second term is to remove the monopole contribution to  
$\frac{\delta H_{phys}}{\delta\Pi^{\sigma}(x)}$. To see this, consider that, by construction,
the monopole contribution to the Lagrangian and Hamiltonian is dependent on the 
time derivatives of $\theta,\phi$, so the monopole component of 
$\frac{\delta H_{phys}}{\delta\Pi^{\sigma}(x)}$ is
\begin{align}
\frac{\delta}{\delta\Pi^{\sigma}(x)} H_{phys} |_{\dot{\theta}\dot{\phi}} 
=& \frac{\delta}{\delta\Pi^{\sigma}(x)} \Big(
\frac{\delta H_{phys}}{\delta \partial_0 \theta(x)} \partial_0 \theta(x)
+ \frac{\delta H_{phys}}{\delta \partial_0 \phi(x)} \partial_0 \phi(x)
\Big) \nonumber \\
=& \frac{\delta}{\delta\Pi^{\sigma}(x)} \Big(
\frac{\delta L_{phys}}{\delta \partial_0 \theta(x)} \partial_0 \theta(x)
+ \frac{\delta L_{phys}}{\delta \partial_0 \phi(x)} \partial_0 \phi(x)
\Big) \nonumber \\
=& \frac{\delta}{\delta\Pi^{\sigma}(x)} \Big(
B_\theta (x) \partial_0 \theta (x) + B_\phi (x) \partial_0 \phi(x) \Big),
\end{align}
which is a consistency condition for eq.~(\ref{eq:altEOMc}). This confirms the necessity of
treating the monopole as a non-dynamic field.

We now observe that
\begin{equation}
\frac{\delta B_{\theta}(x)}{\delta c^{\sigma}(y)} = \frac{\delta B_{\phi}%
(x)}{\delta c^{\sigma}(y)} = 0,
\end{equation}
from which it follows that the EOM of
$c_{\sigma}$ receives no correction. However its $\{,\}_{phys}$ contribution,
corresponding to the terms in the conventional EOM for the Abelian component,
already contains a contribution from the monopole field strength.

Repeating the above steps for the valence gluons $\vec{X}_{\mu}$, assuming $\sigma \ne 0$
and combining
\begin{equation}
\hat{D}_{0}\vec{\Pi}_{\sigma}(x) =\frac{\delta H}{\delta\vec{X}^{\sigma}(x)}
-\int dy^{4} \sum_{\alpha=\phi,\theta} 
\frac{\delta B_{\alpha}(y)}{\delta\vec{X}^{\sigma}(x)}\partial^{0}\alpha(y).
\end{equation}
with
\begin{equation}
\frac{\delta B_{\phi}(y)}{\delta\vec{X}^{\sigma}(x)} =-\Big(\Big(
\sin\phi(y)_{y}\partial^{\sigma}\theta(y)\hat{n}
+\hat{n}_{\theta}(y)\times\vec{X}_{\sigma}(y)\Big) 
\times \vec{X}_0 
-\hat{n}_\phi(y) \hat{n} \cdot \vec{F}_{0\sigma} \Big)
\end{equation}%
\begin{equation}
\frac{\delta B_{\theta}(y)}{\delta\vec{X}^{\sigma}(x)} =\Big(\Big(
\partial_{\sigma}\phi(x)\hat{n}+\sin\phi(x)\, 
\hat{n}_{\phi}(x)\times\vec{X}_{\sigma}(x)\Big) 
\times \vec{X}_0
-\sin\phi \hat{n}_{\theta}(y) \hat{n} \cdot \vec{F}_{0\sigma}\Big) \delta^{4}(x-y),
\end{equation}
gives
\begin{align}
\hat{D}_{0}\vec{\Pi}_{\sigma}(x) =  &  
\frac{\delta H}{\delta\vec{X}^{\sigma}(x)} 
-\frac{1}{2}\Big(\Big(\sin\phi(x)(\partial_{\sigma}\phi(x)\partial_{0}%
\theta(x) -\partial_{\sigma}\theta(x)\partial_{0}\phi(x)\Big)\hat{n}(x)\nonumber\\
& +\Big(\sin\phi(x)\hat{n}_{\phi}(x)\partial_{0}\theta(x) -\hat{n}_{\theta}%
(x)\partial_{0}\phi(x)\Big) \times\vec{X}_{\sigma}(x)\Big)\times \vec{X}_0 \nonumber\\
=  &  \frac{\delta H}{\delta\vec{X}^{\sigma}(x)} -\frac{1}{2}
g^{2}\Big(\vec{C}_{\sigma}(x)\times\vec{C}_{0}(x) +\vec{C}_{0}(x)\times
\vec{X}_{\sigma}(x) \Big)\times\vec{X}_{0}(x).\label{eq:XEOM}%
\end{align}
This is the converse situation of the Abelian gluon, where 
their derivatives $\vec{X}_{\sigma}$
is uncorrected while their EOM
receives a correction which cancels the monopole's electric contribution to
$\{\hat{D}_{0}\vec{X}_{\sigma},H_{phys}\}_{phys}$. This is required by
the conservation of topological current.

\section{The fundamental representation}

\label{sec:fundamental} We consider a complex boson field $\mathbf{a}%
(x),\mathbf{a}^{\dagger}(x)$ in the fundamental representation of the gauge
group, and probe the implications of this approach for the quark fields.
Although physical quarks are fermions, we study the bosonic case to avoid
distracting complications, leaving the fermionic case for a later paper.

The kinetic and interaction terms are given by
\begin{equation}
-(\hat{D}^{\mu}\mathbf{a})^{\dagger}(x) \hat{D}_{\mu}\mathbf{a}(x)
\end{equation}
We do not consider the mass term which makes no contribution to the physics
considered here.

The contribution of $\mathbf{a}(x),\mathbf{a}^{\dagger}(x)$ to $B_{\phi
}(x),B_{\theta}(x)$ is
\begin{align}
B_{\phi}(x)_{|\mathbf{a},\mathbf{a}^{\dagger}} =  &  (\hat{D}^{0}
\mathbf{a}(x))^{\dagger}\hat{n}_{\theta}(x) \mathbf{a}(x) + (\hat{n}_{\theta
}(x) \mathbf{a}(x))^{\dagger}\hat{D}^{0} \mathbf{a}(x)\nonumber\\
B_{\theta}(x)_{|\mathbf{a},\mathbf{a}^{\dagger}} =  &  -(\hat{D}^{0}
\mathbf{a}(x))^{\dagger}\sin\phi(x) \hat{n}_{\phi}(x) \mathbf{a}(x) -(\sin
\phi(x) \hat{n}_{\phi}(x) \mathbf{a}(x))^{\dagger}\hat{D}^{0} \mathbf{a}(x)
\end{align}

leading to a contribution of
\begin{align}
\label{eq:fundcurvature}\mathcal{F}_{\theta\phi}(x)_{|\mathbf{a}%
,\mathbf{a}^{\dagger}} =  &  -(\hat{D}_{0} \mathbf{a}(x))^{\dagger}(\cos
\phi(x) \hat{n}_{\phi}(x)- \sin\phi(x) \hat{n}(x)) \mathbf{a}(x)\nonumber\\
& -(\partial_{0} \theta(x) (\cos\phi(x) \hat{n}_{\phi}(x) - \sin\phi(x)
\hat{n}(x))\mathbf{a}(x))^{\dagger}\sin\phi(x) \hat{n}_{\phi}(x)
\mathbf{a}\nonumber\\
& -((\cos\phi(x) \hat{n}_{\phi}(x) - \sin\phi(x) \hat{n}(x))\mathbf{a}%
)^{\dagger}\hat{D}_{0} \mathbf{a} (x)\nonumber\\
& - (\sin\phi(x) \hat{n}_{\phi}(x) \mathbf{a}(x))^{\dagger}
(\cos\phi(x) \hat{n}_{\phi}(x) - \sin\phi(x) \hat{n}(x))\partial_{0} \theta(x) \mathbf{a}\nonumber\\
&  + (\hat{n}_{\theta\theta} (x) \mathbf{a}(x) \partial_{0} \phi(x))^{\dagger
}\hat{n}_{\theta}(x) \mathbf{a}(x)\nonumber\\
& + (\hat{n}_{\theta}(x) \mathbf{a}(x))^{\dagger}\hat{n}_{\theta\theta}(x)
\mathbf{a}(x) \partial_{0} \phi(x)\nonumber\\
& - (\hat{D}_{0} \mathbf{a}(x))^{\dagger}\hat{n}_{\theta\theta}(x)
\mathbf{a}(x) - (\hat{n}_{\theta\theta}(x) \mathbf{a}(x))^{\dagger}\hat{D}_{0}
\mathbf{a}(x)
\end{align}
to the $q^{\alpha}$-curvature. It follows that the complete expression for the
$q^{\alpha}$-curvature in this theory is the sum of 
eqs.~(\ref{eq:curvature},\ref{eq:fundcurvature})

As with the gluon DOFs, the non-zero $\mathcal{F}_{\theta\phi}(x)$ leads to
the cancellation of the monopole interactions, and generates corrections to
the canonical commutation relations.

\section{Monopole corrections to the quantum commutation relations}

\label{sec:commutator}

Corrections to the classical Poisson bracket correspond to corrections to the
equal-time commutators in the quantum regime. Denoting conventional
commutators as $[,]_{phys}$ and the corrected ones as $[,]_{new}$, for
$\mu,\nu\ne0$ we have
\begin{align} \label{eq:ccnew}
[c_{\mu}(x),c_{\nu}(z)]_{new} = &  [c_{\mu}(x),c_{\nu}(z)]_{phys}
 -\int dy^{4}\Big( \frac{\delta B_{\theta}(y)}{\delta\Pi^{\mu}(x)}
\mathcal{F}_{\theta\phi}^{-1}(z)
\frac{\delta B_{\phi}(y)}{\delta\Pi^{\nu}(z)} - \frac{\delta B_{\phi}%
(y)}{\delta\Pi^{\mu}(x)} \mathcal{F}_{\phi\theta}^{-1}(z)
\frac{\delta B_{\theta}(y)}{\delta\Pi^{\nu}(z)}
\Big)  \delta^{4}(x-z)\nonumber\\
=  &  [c_{\mu}(x),c_{\nu}(z)]_{phys} \nonumber\\
&- \sin \phi(x) \sin\phi(z)( \partial_{\mu}%
\phi(x)\partial_{\nu}\theta(z)- \partial_{\nu}\phi(z)\partial_{\mu}\theta(x)) 
\mathcal{F}_{\theta\phi}^{-1}(z) \delta^{4}(x-z).
\end{align}
The second term on the final line, after integration over $d^4 z$,
clearly becomes
\begin{align}
H_{\mu\nu} (x) \sin\phi(x) \mathcal{F}_{\theta\phi}^{-1}(x),
\end{align}
indicating the role of the monopole condensate in the correction.
By contrast, the commutation relations 
\begin{equation}
[c_{\mu}(x),\Pi_{\nu}(z)]_{new} = [c_{\mu}(x),\Pi_{\nu}(z)]_{phys}, \;
[\Pi_{\mu}(x),\Pi_{\nu}(z)]_{new} = [\Pi_{\mu}(x),\Pi_{\nu}(z)]_{phys} ,
\end{equation}
are unchanged.
Nonetheless, the deviation from the canonical commutation shown in 
eq.~(\ref{eq:ccnew}) is inconsistent with the
particle creation/annihilation operator formalism of conventional second quantization.

Repeating for the valence gluons,
\begin{align} \label{eq:PiXnew}
[\Pi^a_{\mu}(x),\Pi^b_{\nu}(z)]_{new} =  &  [\Pi^a_{\mu}%
(x),\Pi^b_{\nu}(z)]_{phys} -\int dy^{4} \Big( \frac{\delta B_{\theta}%
(y)}{\delta X_a^{\mu}(x)} \frac{\delta B_{\phi}(y)}{\delta X_b^{\nu}(z)}
- \frac{\delta B_{\phi}(y)}{\delta X_a^{\mu}(x)} \frac{\delta B_{\theta
}(y)}{\delta X_b^{\nu}(z)} \Big) \mathcal{F}_{\theta\phi}^{-1}(z)\nonumber\\
=  &  [\Pi^a_{\mu}(x),\Pi^b_{\nu}(z)]_{phys}\nonumber\\
+ &\Big(\sin\phi(z) n^a_\phi(x) n^b_\theta(z) \vec{F}^{0\mu}(x) \cdot \hat{n}(x)
\vec{F}^{0\nu}(z) \cdot \hat{n}(z)
- \sin\phi(x) n^a_\theta(x) n^b_\phi(z) \vec{F}^{0\mu}(z) \cdot \hat{n}(z)
\vec{F}^{0\nu}(x) \cdot \hat{n}(x) \Big) \nonumber \\
&\times \mathcal{F}_{\theta\phi}^{-1}(z)
\delta^{4}(x-z),
\end{align}
where the second term on the final line, integrates over $d^4 z$ to become
\begin{align}
(n^a_\phi(x) n^b_\theta(x) - n^a_\theta(x) n^b_\phi(x))\sin\phi(x)
\vec{F}^{0\mu}(x) \cdot \hat{n}(x) \vec{F}^{0\nu}(x) \cdot \hat{n}(x)
\mathcal{F}_{\theta\phi}^{-1}(x),
\end{align}
while
\begin{equation}
[X^a_{\mu}(x),\Pi^b_{\nu}(z)]_{new} = [X^a_{\mu}(x),\Pi^b_{\nu}(z)]_{phys}, \; 
[X^a_{\mu}(x),X^b_{\nu}(z)]_{new} = [X^a_{\mu}(x),X^b_{\nu}(z)]_{phys} .
\end{equation}

Indeed, this is not an exhaustive presentation of deviations from canonical
quantisation. If a $q^{\alpha}$-gauge field's derivative with respect to any
physical field or its conjugate momentum is non-zero, then that field's
quantisation conditions and particle interpretation are affected unless the
$q^{\alpha}$-curvature is exactly zero. Hence any field interacting with the
monopole component ceases to have a particle interpretation in the presence of
the monopole component. In particular, its particle number becomes
ill-defined, which is reminiscent 
of the parton model.

Eq. (\ref{eq:ccnew}) has a superficial similarity to Dirac brackets. The difference
between our new brackets $\{,\}_{new}$
and Dirac brackets is clarified in app.~B of \cite{D10}. If one introduces
additional "nonphysical" momenta $p_\alpha$ (equation (B1) in
\cite{D10} or sec.~5 of \cite{D14})
corresponding to the "nonphysical" coordinates $q_\alpha$, then the new bracket in
the fully extended phase space becomes the Dirac bracket. But then we
obtain constraints, especially the complicated second stage constraint
equations (B5) (of \cite{D14}), which are absent in our approach. 
Eq.~(\ref{eq:PiXnew}) can therefore be
considered to be a new shortened version of quantization
for singular systems, as described in the conclusions of \cite{D10} and \cite{D14}.

Arguments that coloured states are ill-defined in the infrared regime, based on either
unitarity and/or gauge invariance \cite{O78,OH79,KO79} 
date back several decades but, to our 
knowledge, we are the first to argue that canonical quantisation breaks down.

\section{Discussion}

\label{sec:discuss} We have applied the Clairaut-type formalism to the CDG
decomposition. This has shed light on the dynamics of the topologically
generated chromomonopole field of QCD. In particular, it addresses the issue
of its EOMs, or lack thereof \cite{CP02,BCK02}, 
and the contribution its DOFs
make to the evolution of other fields.

Indeed, the $q^{\alpha}$-curvature was found to be non-zero, leading to
corrections to the time derivatives of the gluon's dynamic DOFs, which
cancel all interactions between physical and non-physical fields from the 
EOMs. This is both necessary for the consistency of eq.~(\ref{eq:altEOMc}),
and qualitatively
consistent with our later finding that the chromomonopole background
alters the canonical commutation relations in such a way as to invalidate 
the particle interpetation of the physical DOFs.

This can be taken to mean that quarks and gluons do not have a well-defined
particle number in the monopole condensate, suggestive of both confinement and
the parton model, but it remains to repeat this work with a fully quantised, 
\emph{i.e.} including ghosts, $SU(3)$ gauge
field, and with fermionic quarks rather than scalar ones. Furthermore, while
many papers have found the monopole condensate \cite{S77,NO78,F80}, especially
with the CDG decomposition \cite{CP00,CP02,CmeP04,Cme04}, to be energetically
favourable to the perturbative vacuum, this result needs to be repeated within
the Clairaut-based quantisation scheme of this paper before strong claims are made.

In summary, this approach offers a rigorous analytic tool for elucidating the
role of topological DOFs in the dynamics of quantum field theories, and finds that 
coloured states have an ill-defined particle number in the presence of non-zero monopole
field strength.
\begin{acknowledgments}
The author S.D. 
is thankful to J. Cuntz and R. Wulkenhaar for kind hospitality 
at the University of M\"unster, where
 the work in its final stage 
was supported by the project ``Groups, Geometry and Actions'' (SFB 878).
\end{acknowledgments}
\appendix

\setcounter{section}{0} \setcounter{equation}{0}

\renewcommand{\thesection}{\Alph{section}}

\section{The Clairaut type formalism}

\label{app:Clairaut}

Here we review the main ideas and formulae of the Clairaut-type formalism for
singular theories \cite{D11,D10}. Let us consider a singular Lagrangian
$L\left(  q^{A},v^{A}\right)  =L^{\mathrm{deg}}\left(  q^{A},v^{A}\right)  $,
$A=1,\ldots n$, which is a function of $2n$ variables ($n$ generalized
coordinates $q^{A}$ and $n$ velocities $v^{A}=\dot{q}^{A}=dq^{A}/dt$) on the
configuration space $\mathsf{T}M$, where $M$ is a smooth manifold, for which
the Hessian's determinant is zero. Therefore, the rank of the Hessian matrix $W_{AB}%
=\tfrac{\partial^{2}L\left(  q^{A},v^{A}\right)  }{\partial v^{B}\partial
v^{C}}$ is $r<n$, and we suppose that $r$ is constant. We can rearrange the
indices of $W_{AB}$ in such a way that a nonsingular minor of rank $r$ appears
in the upper left corner. Then, we represent the index $A$ as follows: if
$A=1,\ldots,r$, we replace $A$ with $i$ (the \textquotedblleft
regular\textquotedblright\ index), and, if $A=r+1,\ldots,n$ we replace $A$
with $\alpha$ (the \textquotedblleft degenerate\textquotedblright\ index).
Obviously, $\det W_{ij}\neq0$, and $\operatorname{rank}W_{ij}=r$. Thus any set
of variables labelled by a single index splits as a disjoint union of two
subsets. We call those subsets regular (having Latin indices) and degenerate
(having Greek indices). As was shown in \cite{D11,D10}, the \textquotedblleft
physical\textquotedblright\ Hamiltonian can be presented in the form%
\begin{equation}
H_{phys}\left(  q^{A},p_{i}\right)  =\sum_{i=1}^{r}p_{i}V^{i}\left(
q^{A},p_{i},v^{\alpha}\right)  +\sum_{\alpha=r+1}^{n}B_{\alpha}\left(
q^{A},p_{i}\right)  v^{\alpha}-L\left(  q^{A},V^{i}\left(  q^{A}%
,p_{i},v^{\alpha}\right)  ,v^{\alpha}\right)  , \label{hph1}%
\end{equation}
where the functions%
\begin{equation}
B_{\alpha}\left(  q^{A},p_{i}\right)  \overset{def}{=}\left.  \dfrac{\partial
L\left(  q^{A},v^{A}\right)  }{\partial v^{\alpha}}\right\vert _{v^{i}%
=V^{i}\left(  q^{A},p_{i},v^{\alpha}\right)  } \label{h}%
\end{equation}
are independent of the unresolved velocities $v^{\alpha}$ since
$\operatorname{rank}W_{AB}=r$. Also, the r.h.s. of (\ref{hph1}) does
not depend on the degenerate velocities $v^{\alpha}$%
\begin{equation}
\dfrac{\partial H_{phys}}{\partial v^{\alpha}}=0, \label{hpv}%
\end{equation}
which justifies the term \textquotedblleft physical\textquotedblright. The
Hamilton-Clairaut system which describes any singular Lagrangian classical
system (satisfying the second order Lagrange equations) has the form%
\begin{align}
  \dfrac{dq^{i}}{dt}=&\left\{  q^{i},H_{phys}\right\}  _{phys}-\sum
_{\beta=r+1}^{n}\left\{  q^{i},B_{\beta}\right\}  _{phys}\dfrac{dq^{\beta}%
}{dt},\ \ i=1,\ldots r\label{q1}\\
  \dfrac{dp_{i}}{dt}=&\left\{  p_{i},H_{phys}\right\}  _{phys}-\sum
_{\beta=r+1}^{n}\left\{  p_{i},B_{\beta}\right\}  _{phys}\dfrac{dq^{\beta}%
}{dt},\ \ i=1,\ldots r\label{q2}\\
\sum_{\beta=r+1}^{n}&\left[  \dfrac{\partial B_{\beta}}{\partial q^{\alpha}%
}-\dfrac{\partial B_{\alpha}}{\partial q^{\beta}}+\left\{  B_{\alpha}%
,B_{\beta}\right\}  _{phys}\right]  \dfrac{dq^{\beta}}{dt}\nonumber\\
&  =\dfrac{\partial H_{phys}}{\partial q^{\alpha}}+\left\{  B_{\alpha
},H_{phys}\right\}  _{phys},\ \ \ \ \ \ \ \ \ \ \ \ \alpha=r+1,\ldots,n
\label{q3}%
\end{align}
where the \textquotedblleft physical\textquotedblright\ Poisson bracket (in
regular variables $q^{i}$, $p_{i}$) is%
\begin{equation}
\left\{  X,Y\right\}  _{phys}=\sum_{i=1}^{n-r}\left(  \frac{\partial
X}{\partial q^{i}}\frac{\partial Y}{\partial p_{i}}-\frac{\partial Y}{\partial
q^{i}}\frac{\partial X}{\partial p_{i}}\right)  . \label{xyp}%
\end{equation}

Whether the variables $B_{\alpha}\left(  q^{A},p_{i}\right)  $ have a
nontrivial effect on the time evolution and commutation relations is
equivalent to whether or not the so-called \textquotedblleft$q^{\alpha}$-field
strength\textquotedblright%
\begin{equation}
\mathcal{F}_{\alpha\beta}=\dfrac{\partial B_{\beta}}{\partial q^{\alpha}%
}-\dfrac{\partial B_{\alpha}}{\partial q^{\beta}}+\left\{  B_{\alpha}%
,B_{\beta}\right\}  _{phys} \label{f}%
\end{equation}
is non-zero. See \cite{D10,D14,D11} for more details.

\end{document}